# Mass spectrometric identification of $C_{60}$'s fragmentation regimes under energetic $Cs^+$ bombardment

Sumaira Zeeshan[a], Sumera Javeed[a] and Shoaib Ahmad[b,*]

[a]Pakistan Institute of Nuclear Science and Technology (PINSTECH), P.O. Nilore, Islamabad, Pakistan

[b]Government College University (GCU), CASP, Church Road, Lahore 54000, Pakistan

[*]National Centre for Physics, Quaid-i-Azam University Campus, Shahdara Valley, Islamabad, 44000, Pakistan

Email: sahmad.ncp@gmail.com

Abstract

Three $C_{60}$ fragmentation regimes in fullerite bombarded by $Cs^+$ are identified as a function of its energy. $C_2$ is the major species sputtered at all energies. For $E(Cs^+) < 1keV$ $C_2$ emissions dominate. $C_2$ and $C_1$ have highest intensities between 1-3 keV with increasing contributions from $C_3$ and $C_4$. Intensities of all fragments maximize around 2 keV. Above 3 keV, fragments' densities stabilize. The roles of and the contributions from direct recoils and collision cascades are determined. Maximum direct recoil energy delivered to the fullerite's $C_{60}$ cage ~210 eV at which only $C_2$ emissions occur is identified and an explanation provided. The three fragmentation regimes under continued $Cs^+$ bombardment eventually lead to complete destruction of the $C_{60}$'s cages transforming fullerite into amorphous carbon.

# 1. Introduction

Various physical aspects and structural properties of fullerenes have been extensively investigated using energetic particle irradiations. The particles range from electrons to ions with energies extending from few hundreds of eV to many MeV [1-9]. Fragmentation of $C_{60}$ has been observed to occur predominantly, in most irradiations, by $C_2$ emission. This is a major fragmentation channel by which $C_{60}$ cage releases its excess energy. This route shrinks the cage by a hexagon. The cage can also fragment irreversibly into smaller components under certain circumstances with $C_1$, $C_3$, $C_4$ accompanying the $C_2$s. Fullerite is the condensed phase of the Buckyball. It has $C_{60}$s arranged on a fcc matrix. Its irradiation with energetic ions leaves traceable structural effects into the solid. Our present experiment utilizes a SNICS II ion source to deliver variable energy $Cs^+$ ions into fullerite to study energetic $Cs^+$ induced fragmentation. The objective of the experiment is to deliver energy in the range ~200 to 5000 eV to fullerite for the purpose of monitoring and to be able to control the nature of the $Cs^+$ induced fragmentation as a function of energy of the bombarding ions. It had earlier been observed that in addition to $C_2$, monatomic C and larger C clusters are among the fragments from $C_{60}$ when bombarded with energetic ions. Varying $Cs^+$ energy between 0.4 to 5.0 keV, we have identified three distinct fragmentation regimes. These vary from $C_2$ only emission when the cage shrinks to the completely fragmented $C_{60}$ cages leading eventually to the complete destruction of the fullerite structure. We also report here the upper threshold energy at which the fragmentation of the $C_{60}$ cages takes place by $C_2$ emission only. That cage energy is found to be ~ 210 eV. At lower than this threshold energy, the cage continues to emit $C_2$s while at higher energies; $C_1$, $C_3$ and $C_4$ provide additional fragmentation routes. $C_2$ however, remains the dominant species sputtered at all energies; a justification for which will be provided in this communication.

SNICS proved to be an ideal arrangement for our recent experiments to study the sputtered C clusters from the condensed $C_{60}$ and graphite targets. The high electron affinities of the C clusters facilitate their detection. The clusters are sputtered predominantly in neutral state and detected subsequently as anions after passing through the neutral Cs° coated target surface and the electron cloud in the ionizer. Detection of anions has its advantages over that of the cations. Production of positive charges requires hot plasmas to remove at least one electron from the respective clusters. Our group's earlier work with the regenerative soot [10] has shown that this process is less efficient and more complex due to the possibility of fragmentation during excitation and differences in ionization mechanisms for a variety of C clusters. SNICS delivers the sputtered species as neutrals in the ground state with one or more electrons attached, while the emitted species from plasma sources are invariably in the excited and ionized states. The cooler environment of SNICS as compared to that of the high temperature graphite discharges is responsible for the interesting results that we have obtained and reported here. Especially, $C_2^+$ is a difficult species to produce and detect as we have reported earlier [11].

## 2. Experimental set-up

The fullerite samples were grown at PINSTECH in Cu bullets by annealing the $C_{60}$ powder at 500˚C. The Cu bullets containing fullerite were used as targets for NEC's SNICS II negative ion source mounted on the 2 MV Pelletron at GCU, Lahore. SNICS provides an excellent experimental set up for removing the target material at a desired rate by sputtering while the bombarding energy can be continuously varied. The source was operated with energy of the $Cs^+$ ions $E(Cs^+)$ between 200 and 5000 eV. The negative carbon atoms and clusters $C_x^-$ ( $x \geq 1$) were extracted from the source at constant beam energy of 30 keV while the target bias was varied between 0.2 to 5.0 keV. A 30 degree bending magnet analyzed the anions.

The experimental method is to start with $E(Cs^+)$ = 5.0 keV, the target surface was sputter-cleaned for 600 sec, and the mass spectrum was obtained from the pristine samples. Figure 1 shows the mass spectra from fullerite 1(a) and graphite in 1(b). Comparison of the fragmentation pattern of the two $sp^2$-bonded allotropes of carbon reveals the nature of the ion induced damage mechanisms in fullerite and graphite. The first major difference is in the intensities of the sputtered species. Emissions from graphite are higher by more than three orders of magnitude compared with those from fullerite. This relates to the nature of the differences in the two allotropic forms of carbon and the respective radiation-induced damage in the two structures. Fullerite has $C_{60}$s arranged on a lattice that is open in the intra-$C_{60}$ region and relatively higher mass concentration in the inter-$C_{60}$ space. Graphite, on the other hand has graphene sheets connected via Van der Waals bonding. Direct recoils- DRs [12] in the case of graphite are composed of $C_1$s only, whereas, in fullerite, $C_{60}$s and $C_1$s act as the primary collision partners (DRs). The nature of the ensuing cascades is different in the two structures. This aspect will be further discussed after presenting the results in Figs. 2 to 5.

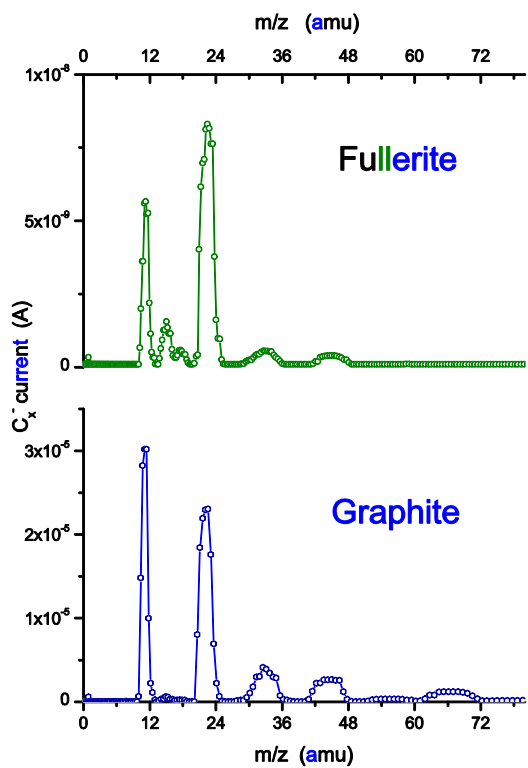

Figure 1

**Fig. 1.** Mass spectra from $Cs^+$ bombarded fullerite and graphite are shown for $E(Cs^+) = 5$ keV. Fullerite and graphite are embedded in Cu bullets as targets for SNICS II source. The two spectra show the carbon atoms and clusters $C_x^-$; x=1 to 4 in the case of fullerite and from 1 to 8 for graphite. There is a huge difference in the cluster intensities between those from graphite ~ $10^{-5}$ A and the ones sputtered from fullerite ~$10^{-9}$ A.

In Fig. 1(a) the $C_2$ peak is higher than that of $C_1$ while the opposite is true for the graphite sample. The other important difference is the recognizable presence of $C_5$, $C_6$, $C_7$, $C_8$ and higher clusters emitted from graphite, while fullerite shows only traces of $C_5$ and $C_6$. The $O^-$ and $OH^-$ impurity at m/z =15 and 16, respectively, is present in both the spectra, indicating the not-so-easily removable water traces.

# 3. Results

## 3.1. Fragmentation pattern from $Cs^+$ variations

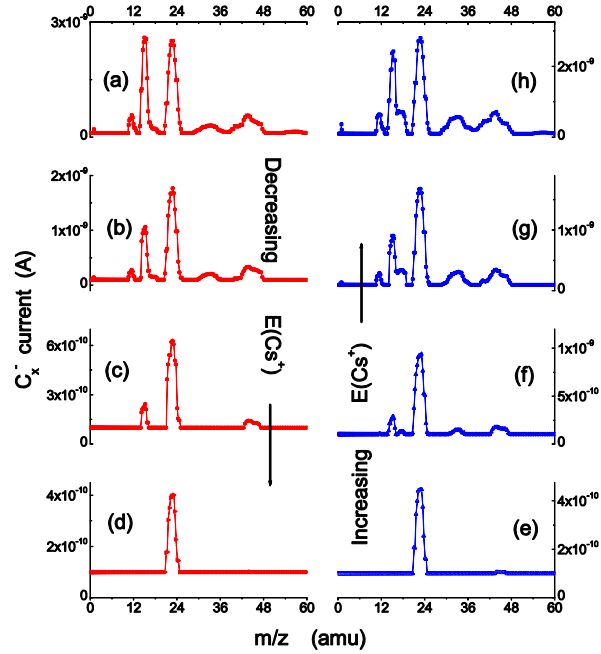

**Fig. 2.** Eight spectra of the intensities of the negative ions $C_x^-$ ($x \geq 1$) from 2(a) to 2(h) are plotted against the m/z. The spectra presented are at four $Cs^+$ energies, 1.0, 0.8, 0.6 and 0.4 keV, respectively. 2(a) to 2(d) are for the decreasing energy while those for the increasing $E(Cs^+)$ are shown from bottom upwards in 2(e) to 2(h). Fig. 2(a) shows two large peaks one due to $C_2^-$ while the other is that of $O^-$. Smaller and broad peaks of $C_1^-$, $C_3^-$ and $C_4^-$ are also present. Reducing $E(Cs^+)$ removes $C_1^-$, $C_3^-$ and $O^-$ completely. At 0.4 keV the only surviving sputtered species is $C_2^-$.

Figure 2

The experiment was conducted in the $Cs^+$ energy range starting at 5000 eV down to 200 eV but the results shown in Fig. 2, for the sake of clarity, are from $E(Cs^+)$=1000 eV to 400 eV in four steps shown in Fig. 2(a) to Fig. 2(d) and then an increasing energy sequence Fig. 2(e) to 2(h). The spectra contain varying intensities of $C_1^-$, $C_2^-$, $C_3^-$, $C_4^-$ anions as a function of the bombarding $Cs^+$ energy. In addition, $O^-$ was seen to be emitted from the $Cs^+$ bombarded fullerite surfaces. In Figure 2 the spectra at the same $E(Cs^+)$ opposite to each other are shown for comparison purposes. $E(Cs^+)$ is lowered towards 0.4 keV where all fragments except $C_2^-$ have vanished. At $E(Cs^+)$ =0.4 keV, only $C_2$ is emitted by the $C_{60}$ cages on the fullerite lattice. The energy range of the spectra shown here is between 1.0 and 0.4 keV to unambiguously illustrate the emergence of the $C_2$ at lower $E(Cs^+)$ as the only fragment. The experiment and the data presented in later figures are for the entire range $0.4 \leq E(Cs^+) \leq 5.0$ keV.

All the spectra shown in Fig. 2 were recorded at 300-500 sec intervals between successive mass analyses. The subtle differences in the two set of spectra i.e., Fig. 2(e) to 2(h) at the same $Cs^+$ energy as shown in 2(d), 2(c), 2(b) and 2(a), respectively, are important for the recognition of fragmentation at $E(Cs^+) > 0.4$ keV. At $E(Cs^+) = 0.4$ keV in Fig. 2(d) and 2(e) recorded successively at 500 sec intervals, one can identify the signatures of the $C_{60}$ cage opening upon receiving ~ 210 eV as a DR in direct collisions with 400 eV $Cs^+$ ion by the process $C_{60} \rightarrow C_{58} + C_2$. The detailed comparisons of the decreasing and increasing $Cs^+$ energy sequences show gross similarities and minute, yet important, differences in the intensities of the fragments. The ubiquitous $C_2$, however, retains its dominance in $Cs^+$ bombardments with decreasing as well as the increasing energy.

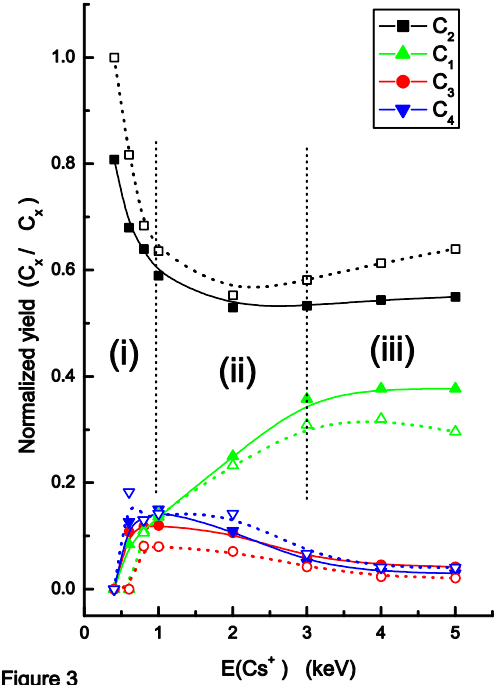

Figure 3

*3.2. Three fragmentation regimes of $C_{60}$*

The dynamics of the fragmentation processes after receiving substantial energy depends upon the amount of energy above the threshold for $C_2$ emission. The process that accompanies cage opening, spitting out $C_2$ in one step or more can be understood if one plots the normalized intensities of the emitted species as is done in Fig.3. The figure plots $C_x/\Sigma C_x$ ; x=1 to 4, for each of the four C species in the form of two sets of data; one each for the reducing and the increasing $E(Cs^+)$. The data shown are from the spectra for the extended energy range i.e. from 0.4 to 5.0 keV. The normalized yields of $C_1^-$, $C_2^-$, $C_3^-$ and $C_4^-$ from the spectra with respective sputtering energies plot and depict the entire landscape of the fragments of the bombarded cages. It can be seen that $C_2^-$ emission dominates over all the fragmentation species at $E(Cs^+) < 1$ keV. At 0.4 keV, the cage receives energy that is enough to force the cage opening for $C_2$ emissions. In the <1 keV energy regime, the relative $C_2^-$ yield decreases sharply with increasing $E(Cs^+)$ while those of $C_1^-$, $C_3^-$, $C_4^-$ increase at slow rates. Above 1 keV, $C_1^-$ competes favorably with $C_2^-$ as a fragment. The relative yields of the two fragments are comparable in energy range from 1.0 to 5.0 keV. The $C_3$ and $C_4$ yields, on the other hand rise gradually with the increasing bombarding energy. Their normalized yields rise initially, go through a broad peak between 1 and 2 keV and has a steady rate of emission ~0.1 of the normalized yield up to 5.0 keV.

*3.3. Fragment intensities during the respective regimes*

The actual cluster intensities are plotted as a function of the $Cs^+$ energy in Fig. 4. It shows total anionic currents. The decreasing and increasing bombarding energy sets show maxima in the intensities of all the species around 2.0 keV. The data suggest that (a) there is a maximum output, and hence the associated cross sections, for the production of all types of fragments around 2 keV, and (b) there is a separation between the region dominated by destruction of the cages and that of the cages opening and closing after emitting $C_2$s. During the initial sequence of the experiment the $Cs^+$ energy was reduced in steps starting from 5.0 keV, until $C_2$ emerged as the only emitted species. The experiment has identified the maximum energy delivered by a $Cs^+$ ion to $C_{60}$ that opens the cage, $C_2$ emitted via $C_{60} \rightarrow C_{58}+C_2$, without destroying the cage irreversibly into smaller components. Reducing the $Cs^+$ energy from 5 keV to 0.2 keV the variations in the intensities and types of the respective spectra's sputtered carbon species ($C_x^-$, x≥1) from the fullerite were recorded. At $Cs^+$ energies >1 keV larger C clusters and monatomic C are emitted along with the dominant $C_2$. As the bombarding energy is decreased below 1 keV, one starts to see the diminishing current densities of all the species except that of $C_2$. At $E(Cs^+) = 0.4$ keV $C_2$ is the only species that is emitted from the fullerite sample, albeit with lower intensity compared with its own emissions at higher $E(Cs^+)$. At and around 400 eV the $C_{60}$ cages respond by opening, spitting out $C_2$ and close again. At this $Cs^+$ energy, the cages may receive up to a maximum of 210 eV. This is the maximum energy transferred in a binary, head-on collision to a massive target like $C_{60}$ by $Cs^+$ as given by Bohr [13] as $E(C_{60})=kE(Cs)$, where $k=4m_1m_2/(m_1+m_2)^2$; $m_1$ and $m_2$ being the masses of Cs and $C_{60}$, respectively. $E(Cs^+)$ was further lowered to 200 eV at which a maximum of ~105 eV is delivered to the cage. $C_2$ being the only emitted species. We could not go any lower in energy as the emitted intensity of $C_2$ is lower than our detection limit of 100 pA. Therefore, the lower energy threshold at which $C_2$ emission stops could not be investigated.

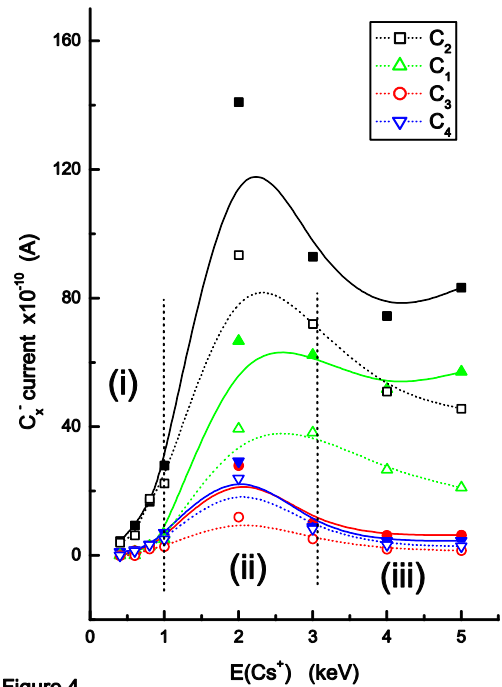

Figure 4

*3.4. Fragment regimes leading to amorphous carbon*

The net effect of the cumulative $Cs^+$ irradiation is the eventual destruction of the cages' structure in fullerite matrix. Fig. 5(a) shows the XRD spectra of the quasi amorphous structure where new broad peaks for 2θ between 15 to 25 degrees identify smaller structure emerging in place of the cages. These are at much reduced intensities. This is in place of the $C_{60}$ in fullerite as shown in Fig. 5(b) with its main peaks at 10.75˚, 17.5˚ and 20.75˚ with smaller ones at higher angles of diffraction. $Cs^+$ being an efficient sputtering agent removes an extensive depth ~ μm at 5 keV. Sputtering of the fragmented components is augmented further by considerable amount of Cs implantation. A carbon target with its much smaller mass than the bombarding $Cs^+$ ion hardly deflects it from its trajectory. This leads to Cs implantation at a depth of ~ 10 nm from the inward receding surface thus saturating the irradiated fullerite. Typically $10^{17-18}$ Cs atoms got implanted during the 8-10 hour experiment. It plays important role in the sputtering profile of the fragments and the subsequent structural transformation from fullerite to a quasi amorphous structure as shown in fig. 5(a).

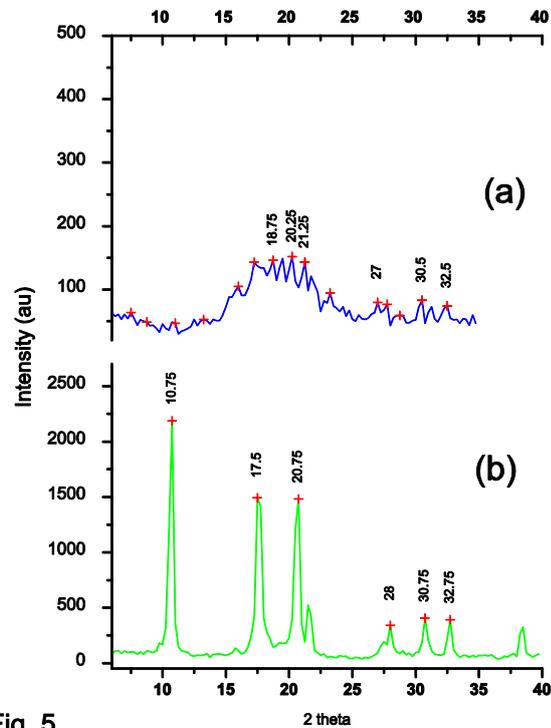

Fig. 5.

# 4. Discussion

The respective contributions of the effects of the Direct Recoil emissions compared with those with the collision cascades [12] can provide an insight into the fragmentation mechanisms. These two closely related aspects of the collisions of the ions with the target constituents in the case of graphite and fullerite are displayed and the resulting nature of the damage illustrated. The first aspect of the ion-target atom collisions is the creation of the primary $C_1$s as DRs. In the case of fullerite or graphite DRs receive energies that depend upon the collision cross section σ(E) that is a function of E, mass ratio $m_2/m_1$ and the impact parameter. The Kinchin and Pease damage function $v_{KP}=E_{DR}/2E_d$ [14] provides a quantitative measure of the damage as the number of displaced atoms. $E_d$ is the threshold energy to displace a target atom permanently. $2E_d$ is needed for the spread of this damage with the creation and spreading of the cascade. $E_{DR}$ is the direct recoil energy of the recoiling $C_1$ in graphite as well as fullerite. Since a $Cs^+$ ion can create ~3 DRs with the maximum energy, the total number of the defects produced is three times larger. For example, 16.3 displaced atoms (or vacancies) per $Cs^+$ are shown to be produced at 1000 eV. This value compares favourably with that obtained from SRIM [15], i.e., 16.7 vacancies/Cs ion at the same energy. The coincidence is due to the fact that SRIM also uses the same definition of the damage function $v_{KP}$. DRs have direct bearing on the spreading of the damage via the collision cascades. The energy density Φ(E) of the recoiling C atoms has $1/E^2$ dependence as a function of the $E(C_1)$ [12]. This favour higher number densities of the displaced atoms with lower energies and is peaked around half the binding energy $E_b$ of C atoms in fullerenes. This is the first and fundamental aspect of the sharing of the incident ion's energy with the target atoms through collision cascades that creates vacancies and interstitials. In graphite, these defects aggregate on or near the surface to produce C clusters $C_x$: x≥1 with the decreasing probabilities for the larger ones. These clusters are detected as anions on being sputtered.

The second aspect of the energy transfer by the incident ion to the target constituents becomes significant in the case of fullerite. Fullerite solids are composed of the $C_{60}$s while graphite's basic unit is $C_1$. C atoms of the individual $C_{60}$ in fullerite have the same cross sectional dependence in collisions with the incident ions as those in graphite and the DRs that are produced. However, at certain impact parameters, the $C_{60}$ cages may receive such amounts of energy which may be just sufficient to fragment it by the emission of $C_2$. That way the excess energy is released. Such energies may be identified as the maximum threshold energy for $C_{60}$ denoted by $E_{DR}(C_{60})$. In this case, we have seen in these experiments that the cage may shrink by $C_{60} \rightarrow C_{58}+C_2$ but without completely fragmenting into smaller components. One can calculate these recoil angles [16] $\Theta_{lab}=\cos^{-1}\{E(C_{60})/E(Cs^+)\}^{1/2}$ at which it receives 210 eV. The corresponding scattering angles of the $Cs^+$ ion $\Psi_{scat}=\sin^{-1}\{\sin \Theta_{lab} (m_2E_2/m_1E_1)^{1/2}\}$ in $Cs^+$-$C_{60}$ collisions that occur to deliver that amount of energy to the $C_{60}$ cage at which it only spits out a $C_2$. For $E(Cs^+) \geq 1000$ eV, this recoil angle $\geq 60°$. It helps to understand that while $C_{60}$ recoils as a DR with low energies, the $Cs^+$ ion also changes its course substantially. At $E(Cs^+)=5000$ eV, the $Cs^+$ ion interacts by scattering at 28° to deliver 210 eV to $C_{60}$ that recoils at 74° from the axis of collision of $Cs^+$-$C_{60}$. This is our proposed explanation of the mechanism by which energy~210 eV can be delivered to $C_{60}$ in direct collisions. At higher $Cs^+$ energies, there are a lot more of these collisions and hence a higher yield of $C_2$s from the fullerite as compared with that from graphite.

## 5. Conclusions

In conclusion one can say that the bombardment of fullerite by heavy energetic $Cs^+$ induces fragmentation of the $C_{60}$ cages by sputtering where the type and number densities of the carbon cluster output can be controlled as a function of the $Cs^+$ energy. The cages fragment in three distinct energy regimes of the bombarding $Cs^+$ ion; in the 0.4 to 1 keV energy regime, $C_2^-$ is the most significant fragment whose maximum normalized yield reduces sharply with increasing the $Cs^+$ energy from 0.4 keV toward 1 keV yet its actual current density continues to grow with all other species. $C_1^-$ is the fragment which is not present at 0.4 keV but its relative normalized yield rises sharply with the falling normalized yield of $C_2^-$. $C_1^-$ characterizes increasing contributions with direct recoils at higher $E(Cs^+)$. $C_3^-$ and $C_4^-$ are less than 10% of all emissions in this energy range. The second energy range with the associated fragmentation regime, is $1 \text{ keV} \leq E(Cs^+) \leq 3$ keV. This regime is the most effective in the production of $C_{60}$ DRs as well as the cage destruction through the collision cascades that lead to higher yields of all the types of fragments $C_1$, $C_2$, $C_3$, $C_4$… These have their intensity maxima within this range around 2 keV. The third energy regime is identified above 3 keV, where the normalized as well as the actual number densities of all fragments stabilize around mean values. The most significant observation of the present investigation is the existence of the maximum $Cs^+$ energy of 400 eV at which $C_2$ is the only fragment observed. This indicates ~210 eV delivered in direct collision by $Cs^+$ to $C_{60}$ in the bombarded fullerite to emit $C_2$ while retaining the cage's integrity by shrinking to a smaller one. At extended irradiation times and doses at $\geq 5$ keV, complete fragmentation of the cages' structure occurs yielding quasi-amorphous, low density carbon. The three well defined fragmentation regimes of $C_{60}$ in fullerite lead to the complete destruction as a function of the bombarding $Cs^+$ energy and dose.

## Acknowledgements:

Authors express their gratitude to Higher Education Commission (HEC), Islamabad for providing substantial funds and grants for the establishment of the 2 MV Pelletron at Government College University (GCU), Lahore. We would like to express our thanks to Rizwan Ali, at PINSTECH, Islamabad with help in preparing the Fullerite samples, and to M. Khaleel at the 2MV Pelletron Lab at GCU, Lahore for providing technical support with SNICS.